\documentclass{article}\usepackage{natbib,epsfig,emulateapj}
\newcommand{\beq}{\begin{equation}}
\newcommand{\eeq}{\end{equation}}
\newcommand{\bea}{\begin{eqnarray}}
\newcommand{\eea}{\end{eqnarray}}

\newcommand{\rem}[1]{ }

\bibpunct[,]{(}{)}{;}{a}{}{,}

\begin{document}

\title{Can Cluster Evaporation Explain the Missing Thermal Energy in Galaxy Clusters?}

\author{Mikhail V. Medvedev{}\altaffilmark{1}}
\affil{
Department of Physics and Astronomy, 
University of Kansas, KS 66045} 
\altaffiltext{1}{
also at the Institute for Nuclear Fusion, RRC ``Kurchatov
Institute'', Moscow 123182, Russia}

\begin{abstract}
Resent observations of a number of galaxy clusters using the 
Sunyaev-Zel'dovich effect indicate that about 1/3 of baryonic mass is
missing from the hot intracluster medium (ICM), which is significantly larger 
than the fraction of stars and cool gas, which account for only about 10\%.
Here we address the question whether the remaining $22\pm10\%$ can be 
accounted for by thermal evaporation of gas from clusters. We have found 
that evaporation can occur only from the cluster ``surface'',
$r\sim r_{\rm vir}$, and not 
from it's interior. We evaluated particle diffusion through the magnetized
ICM for several scenarios of ISM turbulence and found that diffusivity 
is suppressed by at least a factor of 100 or more, compared to the Spitzer 
value. Thus, only particles from radii $r\ga0.9r_{\rm vir}$ can evaporate.
Diffusion of particles from inside 
the cluster, $r\la0.9r_{\rm vir}$, takes longer than the Hubble time. 
This lowers the cluster-averaged fraction of the evaporated 
hot gas to few percent or less. However, if
the missing hot component {\it is indeed} due to evaporation, this 
strongly constrains the magnetic field structure in the cluster envelope,
namely either (i) the gas is completely unmagnetized ($B\le10^{-21}$~gauss) 
in the cluster halo or (ii) the magnetic fields in the ICM are rather 
homogeneous and non-turbulent.
\end{abstract}



\section{Introduction}

Observations of the Sunyaev-Zel'dovich effect in galaxy clusters 
directly probe the energy content of the intracluster medium (ICM)
in the model-independent way, and it is complementary to X-ray observations 
(see \citealp{Voit05} and references therein).
The study of 193 clusters with temperatures above 3~keV using the 
cosmic microvawe background data from WMAP yielded a remarkable 
and puzzling result \citep{Afshordi+07}. 
It has been shown that SZ effect would account for 
a half of the thermal energy of the ICM provided all the baryons are in the 
hot phase. This result implies that a significant fraction of baryons, 
about $32\pm10\%$ by mass, is missing from the ICM phase. 

It has been proposed \citep{Loeb07} that gas evaporation can reduce 
internal energy of gas in a cluster.
Evaporation involves particles from the 
thermal tail of the Maxwellian distribution because their 
velocities exceed the escape velocity from a cluster, 
hence they are gravitationally unbound and can leave the cluster. 
Since these particles are suprathermal, 
i.e., their speed exceeds the thermal speed, the overall 
energy content of the cluster will be decreasing with time.
Because the ICM gas is mildly collisional, the transport of these 
particles will be diffusive. However,
in the presence of magnetic fields in the ICM, particle diffusion
is often substantially reduced \citep{RR78,CC98,NM01,MK01,M01,L06}. The 
reduction coefficient $f_B<1$ has been introduced; its value has not 
been accurately evaluated, however. 

Magnetic fields in galaxy clusters are highly turbulent with the 
fluctuating component being of order the mean field, $\delta B\sim B$,
and is tangled on small scales. The magnetic field spectrum
deduced from the rotation measure observations is peaked at about 
$l_B\sim{\rm one - few~kpc}$, which sets the characteristic correlation 
length of the field \citep{VE03,VE05,Ensslin+05}. We evaluate the diffusion
suppression coefficient $f_B=\kappa_{\rm eff}/\kappa_{\rm Sp}$ (where
$\kappa_{\rm eff}$ and $\kappa_{\rm Sp}$ are the effective and Spitzer
diffusion coefficients) in the turbulent magnetized ICM. The obtained 
value turns out to be about 0.01 for a range of typical parameters
of the ICM gas. This value is substantially lower than what have
previously been assumed. We discuss implications of this result
for the cluster properties, dynamics and formation.

\section{Particle diffusion in the cluster halo} 

For simplicity we assume that the ICM gas is hydrogen and neglect 
the effect of helium. We also assume that the magnetic
field in the ICM is highly turbulent and chaotic. Particle 
transport in such a system is strongly influenced by the field
properties. 

A particle in the magnetic field gyrates about a certain field line and
can move freely along it. Thus, on average, the particle closely follow the
field line. A particle can jump through the distance 
$\sim r_L=v_{\rm th}m_p c/eB$ (the Larmor radius) 
to a neighboring field line in particle-particle collisions.
Since the Larmor radius is many orders of magnitude smaller than the 
mean-free-path in the ICM, the cross-field diffusion is greatly suppressed.
For future reference, the Larmor radius, the proton mean-free-path for the
typical conditions in the galaxy cluster halo, $r\sim r_{\rm vir}$, and 
the virial radius are
$
r_L = 2.3\times 10^{-12}\ {\rm Mpc}\ T_{\rm 5keV}^{1/2}\ B_{\rm 1nG}^{-1},
$,\ 
$
\lambda = 1.3\ {\rm Mpc}\ T_{\rm 5keV}^2\ \Delta_{25}^{-1},
$,\
$
r_{\rm vir} = 2.3\ {\rm Mpc}\ T_{\rm 5keV}^{1/2},
$
where $\Delta=n_p/\bar n_p\sim 25$ is the overdensity at the present 
epoch, $z=0$, characterizing the ICM gas density at the virial radius, 
$\bar n_p=2\times10^{-7}~{\rm cm}^{-3}$ is the
mean density in the Universe at $z=0$ \citep{Spergel+06}, 
$\Delta_{25}=\Delta/25$, $T_{\rm 5keV}=T/({\rm 5~keV})$, 
the ICM magnetic field $B_{\rm 1nG}=B/({\rm 10^{-9}~gauss})$. 
The thermal and  Alfv\'en speeds are
$
v_{\rm th}=(3kT/m_p)^{1/2}
=1.2\times10^8\ T_{\rm 5keV}^{1/2}\ {\rm cm}~{\rm s}^{-1},
$,\
$
v_A=B/(4\pi m_p n_p)^{1/2}=9.7\times10^4 
B_{\rm nG} \Delta_{25}^{-1/2}\ {\rm cm~s}^{-1}
$.
Below we consider three different scenarios for the particle transport: 
the case of static chaotic fields, the Alfv\'enic cascade and transport
by large-scale turbulent motions (including the hydrodynamic cascade).
We show that transport in Alfv\'enic cascade \cite{L06} is suppressed by 
some power of the Alfv\'enic Mach number, whereas the static fields and 
the fluid motions yield similar values for the diffusion coefficient 
for typical cluster conditions.  

{\it Static chaotic fields} ---
In a medium with strong magnetic turbulence, the magnetic field lines 
themselves can be chaotic, such that the separation of two field lines
increases exponentially with distance along them, 
$d\sim r_L\exp(l_\|/l_B)$, with $l_B$ being the field correlation length.
 
The transport of a proton along such 
field lines results in effective diffusion coefficient 
is  suppressed compared to the classical Spitzer value
$
\kappa_{\rm Sp}= \lambda v_{\rm th}
=4.8\times10^{32}\ {\rm cm}^2~{\rm s}^{-1}\ 
T_{\rm 5keV}^{5/2}\ \Delta_{25}^{-1}.
$
Note that here we shall not use the ambipolar rate, which is 
a factor of two larger.\footnote{ The ambipolar diffusion coefficient,
$
\kappa_{\rm amb}={(T_e+T_p)\kappa_{{\rm Sp},e}\kappa_{{\rm Sp},p}}/
{(T_p \kappa_{{\rm Sp},e} + T_e \kappa_{{\rm Sp},p})},
$
where $\kappa_{{\rm Sp},e}=\lambda_e v_{{\rm th},e}$ and
$\kappa_{{\rm Sp},p}=\lambda_p v_{{\rm th},p}\ll\kappa_{{\rm Sp},e}$
for $T_e\simeq T_p$, following from plasma quasi-neutrality, is
commonly used for two-component $e^-$-$p$ plasma.
Here, we consider diffusion of a supra-thermal component of the proton
distribution through gas, not the global escape of 
the plasma bulk, hence the Spitzer diffusivity is used. }

If the magnetic field has a well defined, single scale $l_B$, 
the suppression factor due to the field line geometry, i.e., the field 
line tangling alone, is \citep{RR78,CC98}
$
f_{\rm field}\simeq \left[3\ln(l_B/r_L)\right]^{-1}\sim{1}/{60}.
$
Here the factor 1/3 appears because the particle motion is effectively 
one-dimensional (i.e., along a field line), rather than three-dimensional. 
Here we also used the observational value of the typical field 
correlation scale $l_B\sim 1$~kpc \citep{VE03,VE05}.

If $l_B$ ranges through two decades or more in scale, as one can 
expect in the inertial range of strong magnetohydrodynamic (MHD) turbulence,
the suppression due to tangling is weaker \citep{NM01}:
\beq
f_{\rm field}\simeq 1/5.
\label{ft2}
\eeq
In this case, the velisity on the injection scale is $V_L=v_A$. 
For a general case $V_L\not=v_A$, see \citet{L06}.

Another effect that is limiting the diffusion rate is 
the magnetic mirroring \citep{CC98,NM01,MK01}. 
Due to the conservation of the adiabatic invariant of a gyrating particle,
$\mu=(m v_\bot^2/2)/B$, the particle may not be able to penetrate through
the region of strong magnetic field, depending on the particle's pitch 
angle and the field strength. In clusters, the field inhomogeneities are
large, $\delta B \sim B$, therefore the mirroring effect can be significant.

Accurate calculation of the diffusion suppression due to mirroring in 
various regimes (single and multi-scale) and with various models
of turbulence has been done in an excellent work by \citet{MK01}.
The mirroring suppression depends on the ratio $\lambda/l_B$, an it
is strong when $\lambda\gg l_B$, as in the cluster halo.
Note that the mirroring effect is small in the cluster core where 
$\lambda< l_B$, hence \citet{NM01} argued that $f_{\rm mirroring}\la1$.
The results obtained by \citet{MK01} can be approximated quite well as follows:
%
\beq
f_{\rm mirroring}\simeq\min\left\{
\left[{l_B}/(10^2\lambda)\right]^{1/3}
,\ 1\right\}.
\label{fm}
\eeq

The overall diffusion suppression factor is
\beq
f_B=f_{\rm field}\ f_{\rm mirroring}\sim4.3\times10^{-3}.
\label{fB}
\eeq
Here we used $l_B\sim1$~kpc, and we assumed the most favorable
case of multi-scale turbulence with $f_{\rm field}\sim1/5$. Note that 
even if the magnetic fields are correlated on a cluster size scale, 
$l_B\sim1$~Mpc, the suppression is still rather strong: 
$f_B\sim0.05$. Thus, the magnetic field suppression is much stronger
than $f_B\sim1/3$ assumed in \citep{Loeb07}, for which one have obtained 
that about 10\% of the cluster gas will evaporate during the Hubble time.
Simply re-scaling this result to the value of $f_B\sim4\times10^{-3}$ yields
the evaporated fraction of order $\sim0.1\%$. This result is essentially 
independent of the fraction of open field lines crossing $r_{\rm vir}$.

{\it Alfv\'enic turbulence} --- The particle diffusivity in the Alfv\'enic 
MHD cascade (but neglecting the mirroring effect) has been studied 
by \citet{L06}. He has found that to sub-Alfv\'enic turbulence, 
$v_{\rm gas}/v_A\equiv M_A<1$, the diffusion coefficient is suppressed 
by a factor of $M_A^2<1$ compared to the \citet{NM01} result: 
\beq
f_{\rm field}\sim(1/3)M_A^4, \qquad M_A<1,
\eeq
cf. Eq. (8) in \citet{L06,L07}.
The factor 1/3 above is approximate, however it is not very different 
from the factor 1/5 obtained by \citet{NM01}.

For the super-Alfv\'enic turbulence, i.e., gas motions with the Alfv\'en
Mach number $M_A>1$, there are two regimes. In the first one, the particle 
mean free path is large compared to the so-called ``Alfv\'en length'',  
$\lambda>l_A$, where $l_A\approx l_B M_A^{-3}<l_B$, and the
suppression is 
\beq
f_{\rm field}\sim(1/3)(l_A/\lambda)\sim(l_B/\lambda)M_A^{-3}, 
\qquad M_A>1,
\eeq
cf. Eq. (4) of \citet{L06}, and note that $l_B\ll\lambda$.
In the second case, $\lambda<l_A<l_B$, one recovers the $f_{\rm field}\sim1/3$
limit of one-dimensional diffusion. This case, however is not very 
relevant for cluster evaporation, because the mean free path is 
comparable to the system size and is much larger than the outer scale
of turbulence, $\sim1~{\rm Mpc}\gg l_B$.

As in any magnetic turbulence, the conservation of the magnetic moment
results in additional suppression due to particle trapping, as
discussed above, Eqs. (\ref{fm},\ref{fB}). Since the ``field'' part of 
the suppression factor alone is diminished by some power of the Alfv\'enic 
Mach number in both sub- and super-Alfv\'enic regimes, we conclude that 
the static field model puts less stringent limits on thermal transport.

{\it Fluid turbulence} --- Turbulent fluid eddies can also contribute 
to the particle transport. The dynamical diffusion coefficient has been 
also evaluated by \citet{L06}: 
\beq
\kappa_{\rm dyn}\approx \left\{
\begin{array}{ll}
\displaystyle{C_d\ L V_L}, & M_A\ga1; 
\\
\displaystyle{\beta C_d\ L V_L M_A^3}, & M_A<1,
\end{array}
\right. \qquad
\eeq
where $\beta\sim4$ and $C_d\approx1/3$ \citep{L90} (cf., it is not 2/3
as in \citealp{L06} because we consider proton transport only).
Note that transport in sub-Alfv\'enic turbulence is suppressed by $M_A^3$.
Here $L$ is the eddy scale and $V_L$ is the turbulent velocity on this scale.

It is instructive to represent the Spitzer diffusivity as 
\beq
\kappa_{\rm Sp}\approx0.57\ T_{\rm 5keV}^{3/2}\Delta_{25}^{-1}\
(r_{\rm vir} v_{\rm th}),
\eeq
i.e., it is very close to $r_{\rm vir} v_{\rm th}$ for typical cluster 
halo conditions. In the fluid turbulence, the combination $LV_L$ must 
be substantially smaller than $r_{\rm vir} v_{\rm th}$. Indeed, 
if $LV_L\sim r_{\rm vir} v_{\rm th}$, eddies
of $L\sim r_{\rm vir}/2$ and $V_L\sim 2v_{\rm th}$ break the cluster apart:
in virial equilibrium $v_{\rm esc}^2=2v_{\rm th}^2$ and gas parcels
with $V_L>2^{1/2}v_{\rm th}$ are gravitationally unbound.
Some fraction of gas can be lost in such violent events (e.g., mergers),
but this cannot be called a steady-state evaporation.

Numerical simulations of turbulence in clusters, taking into account
decaying turbulence, mergers and wakes \citep{Dolag+05,Sub+06,Vazza+06} 
yield a typical value of $LV_L\sim4.5\times10^4$~kpc~km~s$^{-1} \sim
0.016\ T_{\rm 5keV}^{-1}\ (r_{\rm vir}v_{\rm th})$. This yields
\beq
\kappa_{\rm dyn}/\kappa_{\rm Sp}\simeq0.009\ T_{\rm 5keV}^{-5/2}\Delta_{25},
\eeq
for $M_A>1$, which is comparable to diffusion in static fields. 
Whether and how much the ICM fields will limit particle escape through
$r_{\rm vir}$ has never been studied, but it goes beyond the diffusion
approximation and may require a numerical approach.

{\it Diffusion length and evaporated gas fraction} ---
Considering several scenarios, we found that particle diffusion 
in the magnetized ICM is suppressed by at least a factor of a hundred 
or more, compared to the Spitzer value.  Thus, the static field scenario 
represents an optimistic estimate (lowest suppression) and we used it 
for further analysis.
We now evaluate the diffusion length, through which a proton can
travel during the Hubble time, $t_H=H^{-1}|_{z=0}\simeq4\times10^{17}$~s:
\beq
l_{\rm diff}=\left(\kappa_{\rm eff}t_H\right)^{1/2} \simeq 
0.27\ {\rm Mpc}\ f_{B,0.004}^{1/2}\ T_{\rm 5keV}^{5/4}\ \Delta_{25}^{-1/2},
\eeq
where $f_{B,0.004}=f_B/0.004$.
Note that the diffusion coefficient is inversely proportional to 
the local gas density, which, in turn, is $\propto r^{-3}$ for the 
NFW profile \citep{NFW97}. 
Formally replacing $\Delta$ with $\Delta(r/r_{\rm vir})^{-3}$,
we see that the diffusion from the inner parts of the cluster is strongly 
suppressed. Hence, the highly collisional gas at $r<0.5r_{\rm vir}$ 
cannot replenish the evaporated suprathermal component of the particle
distribution. Because the escaping supra-thermal protons move at speeds
$\ga v_{\rm esc}\sim \eta^{1/2}v_{\rm th}\sim 2^{1/2}v_{\rm th}$,
the diffusion distance above is 
accurate to a factor of order unity. Since most of plasma particles, both
thermal and supra-thermal, are effectively trapped, their residence time 
in between magnetic mirrors can be comparable to the Hubble time. 
Supra-thermal particles with velocity $v=\eta^{1/2}v_{\rm th}$
can experience $N$ collisions per the Hubble time:
$
N\sim t_H/\tau_{\rm coll}
\sim 12\left(\eta\ T_{\rm 5keV}\right)^{-3/2}\Delta_{25}, 
$
where $\tau_{\rm coll}=\lambda(v)/v=3.3\times10^{16}\ {\rm s}\ 
(\eta T_{\rm 5keV})^{3/2}\ \Delta_{25}^{-1}$ is the particle 
collision time, $\eta\sim2.3$ for $v$ being equal to the escape 
velocity, $v_{\rm esc}$, at $r_{\rm vir}$ \citep{Loeb07}.
Thus, there can be significant re-distribution of trapped, $v<v_{\rm esc}$, 
and untrapped, $v>v_{\rm esc}$, components (i.e., an initially
trapped supra-thermal proton may become untrapped and vice versa, 
due to collisions). This may change, perhaps lower, the effective $\eta$. 
In this regime, numerical computation of the 
effective $\eta$ using a self-consistent proton 
distribution accounting for diffusion, trapping and leaking of 
particles is highly desirable.

The diffusion length is small fraction of the cluster size:
\beq
{l_{\rm diff}}/{r_{\rm vir}}\equiv \xi
=0.12\ f_{B,0.004}^{1/2}\ T_{\rm 5keV}^{3/4}\ \Delta_{25}^{-1/2}.
\label{ldiff}
\eeq
Thus, cluster evaporation effectively occurs from a thin layer of
thickness $\sim l_{\rm diff}\sim0.1r_{\rm vir}$. Particles from
radii $r<(r_{\rm vir}-l_{\rm diff})$ cannot escape from the
cluster within the Hubble time.  

The NFW density profile \citep{NFW97} is $n_{\rm gas}(x)=n_0/[x(1+x)^2]$, 
where $x=r/r_s$, $r_s=r_{\rm vir}/c$, $c$ is the concentration parameter,
$n_0$ is the normalization such that $n(x_{\rm vir})=\bar n_p\Delta$
and $x_{\rm vir}=r_{\rm vir}/r_s=c$. The mass profile of the NFW cluster
is $M(x)=M_0[\ln(1+x)-x/(1+x)]$ with $M_0=4\pi r_s^3 n_0 m_p$.
The fraction of the gas mass in the shell of thickness $l_{\rm diff}$
to the total mass of gas in the entire cluster is
\beq
\frac{M_{\rm shell}}{M_{\rm cluster}}=
\frac{\ln\left(\frac{c+1}{c+1-\xi c}\right)-\frac{\xi c}{(c+1)(c+1-\xi c)}}
{\ln(c+1)-c/(c+1)}.
\eeq
The supra-thermal fraction of the proton distribution depends on the
concentration parameter. It has been evaluated to be 
$0.064<\epsilon<0.092$ at $r=r_{\rm vir}$ and for $2<c<8$ \citep{Loeb07}.
Because of trapping of particles in magnetic traps, the particle 
distribution can evolve, therefore the actual value of $\epsilon$
can differ. The fraction the evaporated gas is, therefore,
\beq
\mu\equiv\epsilon\ \frac{M_{\rm shell}}{M_{\rm cluster}}
\simeq\frac{\epsilon\xi c^2/(c+1)^2}{\ln(c+1)-c/(c+1)}\sim 7.2\times10^{-3},
\eeq
where we used that $\xi=0.12$, $c=4$ and the corresponding 
$\epsilon=0.076$. Thus, the evaporation 
fraction of the ICM gas is about 1\%. Figures \ref{f:1} and \ref{f:2} 
represents the evaporated fraction $\mu$ as a function of the ICM 
temperature and the field correlation length, $l_B$, 
for several values of $c$ and $l_B$.

\section{Discussion}

We evaluated the suppression factor of the diffusive transport
of protons in the ICM with magnetic fields in three scenarios: 
(i) static chaotic fields \citep{NM01}, (ii) Alfv\'enic and (iii)
fluid turbulence \citep{L06}. Note that magnetic fields are still
required even in case (iii), otherwise the fluid approximation 
breaks down, because $\lambda\sim r_{\rm vir}$. We considered 
diffusion suppression is due to (i) a purely geometric effect of
turbulence and (ii) the dynamic effect of
particle mirroring and trapping in inhomogeneous fields. 
We have found that particle diffusion is suppressed to at least 
1\% of the Spitzer value, and usually much more. For further analysis
we used the suppression in static fields as an optimistic representative 
value.

The smallness of the effective diffusivity, compared to the Spitzer value
substantially lowers the overall evaporation gas fraction. 
We calculated the diffusion distance for typical cluster parameters 
and found that a particle can diffuse through the distance of 
about $l_{\rm diff}\sim0.1r_{\rm vir}$ within the Hubble time. 
Thus, particles that are at radii 
$r<(r_{\rm vir}-l_{\rm diff})\sim0.9r_{\rm vir}$ can never 
pass through the virial radius and leave the cluster.
Consequently, cluster evaporation occurs from the thin cluster
``skin'' of thickness $\sim0.1r_{\rm vir}\sim200$~kpc.
The fraction of the gas evaporated from this skin can be rather high,
perhaps about 10\% or even more, comparable to the estimate by \citet{Loeb07}.
The exact value shall be obtained from self-consistent simulations
taking into account particle trapping, particle collisions, 
evolution of the distribution function and particle leaking out of 
the cluster.

We make a prediction that the thermal energy deficit shall be the strongest
at the outskirts of the cluster $r\sim r_{\rm vir}$, in the shell of 
few hundred kpc in thickness. Deeper in the cluster, the thermal 
energy deficit shall be negligible. Of course, at radii of few tens kpc,
a cooling flow, if present, may lower the thermal energy as well. 

\begin{figure}
\plottwo{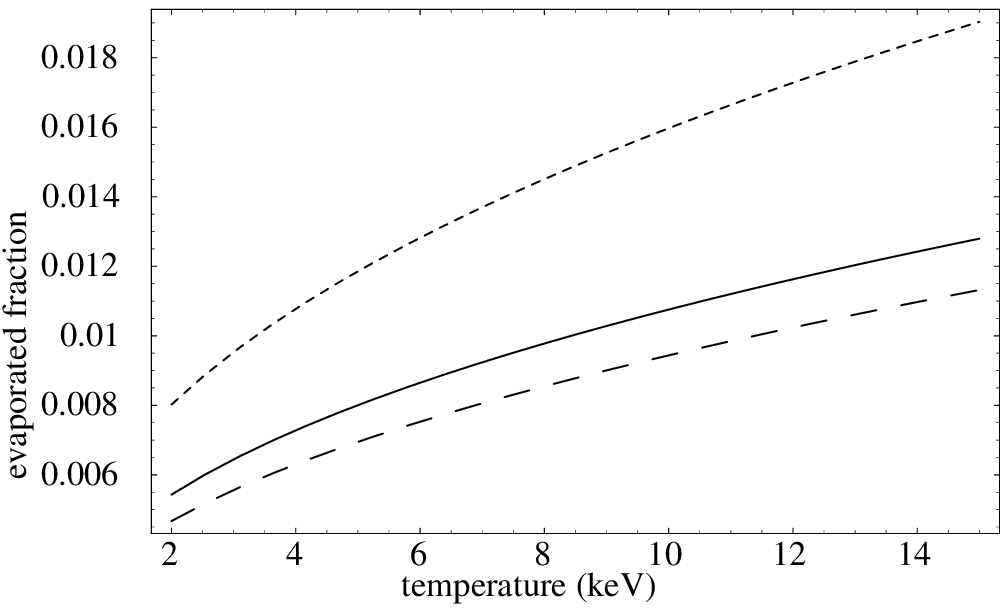}{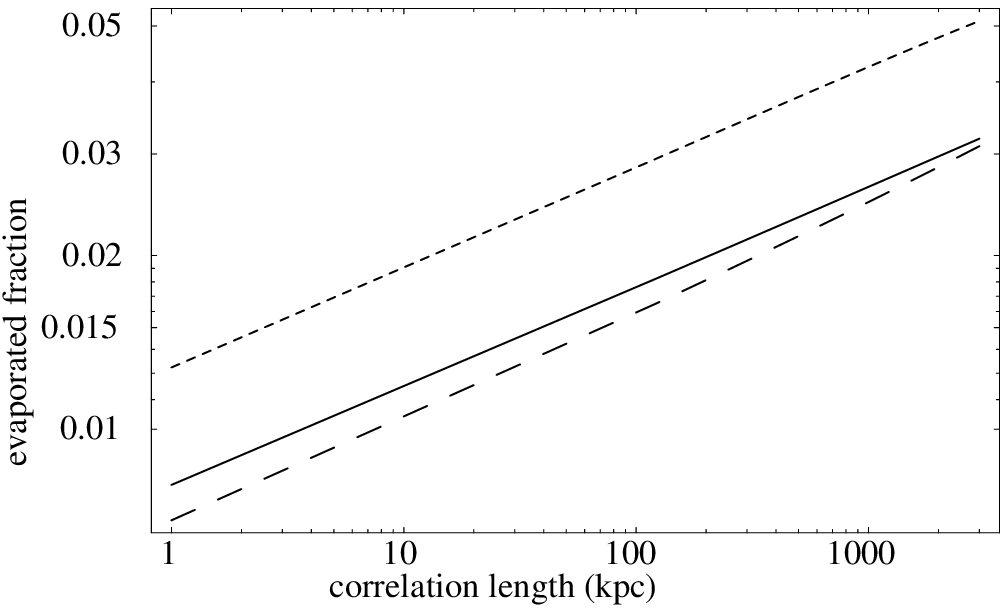}
\caption{Fraction of the ICM gas evaporated from a cluster out of 
$r_{\rm vir}$ as a function of the temperature of the ICM gas. 
The solid curve is computed for $l_B=1$~kpc and $c=2$, 
the dotted is for $l_B=10$~kpc and $c=2$, and the dashed is for 
$l_B=1$~kpc and $c=8$.
\label{f:1} }
\caption{Fraction of the evaporated gas as a function of $l_B$, 
the maximum correlation length the magnetic field.
Here it runs from the observed value of $\sim$1~kpc to the maximum possible 
scale, the cluster size, $\sim$3~Mpc. The solid curve is computed for
$T=5$~keV and $c=2$, the dotted is for $T=15$~keV and $c=2$, and the
dashed is for $T=5$~keV and $c=8$.
\label{f:2} }
\end{figure}

Since the diffusion distance is small compared to the cluster size,
the fraction of the evaporated gas compared to the total cluster mass
is small too. We evaluated it to be less than 1\% for typical cluster 
conditions, $T\sim5$~keV and $l_B\sim1$~kpc. 
Although the concentration parameter $c$ can vary
for different clusters, its effect on the evaporated fraction, $\mu$, 
is not very significant. The ICM gas temperature and the field 
correlation scale do affect $\mu$. Figures \ref{f:1},\ref{f:2}
show how $\mu$ depends on $T$ and $l_B$. Interestingly, by boosting
$l_B$ to $\sim10$~kpc and $T$ to $\sim15$~keV altogether, one can increase
the evaporated fraction to about 2\% only. One cannot exceed the
value of $\mu\sim{\rm few}~\%$ even by pushing $l_B$ to the maximum 
possible value of $\sim3$~Mpc, --- the cluster size. Apparently,
the predicted evaporated gas fraction is below the observational
value of $22\pm10\%$ by a large margin.
 
Now, let's look at the result from a different point of view. 
Let's suppose that
the observed thermal energy deficit of $\sim$20\% {\it is} due to
cluster evaporation. What assumptions shall be relaxed in this case?
We identify two alternatives. First, the ICM shall be effectively 
unmagnetized, then the result of \citet{Loeb07} for the low-suppression 
scenario holds and the evaporated fraction can be of order 10\%.
In order to neglect the effects of magnetic field, the particle
Larmor radius shall be comparable to the size of a cluster, 
$r_L\sim r_{\rm vir}$. This sets the limit on the field strength:
$B\le10^{-21}$~gauss.
Second, if the ICM fields are present, they shall be relatively homogeneous
with $\delta B \ll B$. In this case, particle trapping is small and 
protons can easily escape along open field lines. The evaporated 
fraction will be reduced by the fraction of the cluster surface threaded by
open fields lines. If this suppression is not strong, i.e., of order unity,
then one can again expect $\sim$10\% value for the evaporated gas fraction.
Alternatively, cluster mergers can lead
to some gas loss. However, this process is episodic and is already
taken into account in LSS hydro simulations.

\acknowledgements

We thank the referee, Alex Lazarian, for discussion and comments.
This work has been supported by grants NNG04GM41G, NNX07AJ50G and  
DEFG0204ER54790.

\vskip-1cm

\end{document}